\documentclass[
 aip,
 amsmath,amssymb,
 preprint,
]{revtex4-2}

\usepackage{graphicx}
\usepackage{dcolumn}
\usepackage{bm}

\usepackage[utf8]{inputenc}
\usepackage[T1]{fontenc}
\usepackage{mathptmx}
\usepackage{amsmath}
\usepackage{amssymb}
\usepackage{lipsum}
\usepackage{graphics}
\usepackage[varg]{txfonts}
\usepackage{graphicx}
\usepackage{epsfig}
\usepackage{wrapfig}
\usepackage{rotating}
\usepackage{color}
\usepackage{comment}
\usepackage[normalem]{ulem}
\usepackage{multirow}
\usepackage{tabularx}
\usepackage{graphicx}
\usepackage{dcolumn}
\usepackage{hyperref}

\usepackage{pgfplots}
\usetikzlibrary{pgfplots.groupplots}
\usepackage{tikz}
\usetikzlibrary{positioning}
\usetikzlibrary{calc}
\usepackage{fp}
\usetikzlibrary{fixedpointarithmetic} 

\begin{document}

\title{Modeling Simulated Emissions from Galactic Binary Stars}

\author{Theodora V. Papavasileiou} 
 \email{th.papavasileiou@uowm.gr}
 \affiliation{Department of Informatics, 
University of Western Macedonia, GR-52100 Kastoria, Greece}
\author{Odysseas T. Kosmas}%
 \email{odysseas.kosmas@manchester.ac.uk}
\affiliation{Modelling and Simulation Center, MACE, University of Manchester, Sackville Street, Manchester, UK}

\author{Ioannis Sinatkas}
 \email{isinatkas@uowm.gr}
\affiliation{Department of Informatics, 
University of Western Macedonia, GR-52100 Kastoria, Greece}

\date{\today} 

\begin{abstract}
Relativistic plasma flows from the jets of black hole binary systems consist the environment of multiple particle production and radiation emission including neutrinos and gamma-rays. We implement a hadronic model based on $p-p$ interactions with the purpose of predicting the produced secondary particle distributions inside the jet. Our ultimate goal is the neutrino and gamma-ray intensities calculation while taking into account the most important gamma-ray absorption processes in order to present more realistic results. 
\end{abstract}


\maketitle


\section{Introduction}

Relativistic magnetohydrodynamical jets are plasma flows emanating from the vicinity of black holes 
or neutron stars. They constitute sources of various multi-wavelenth (X-rays, $\gamma$-rays, etc.) 
and multi-particle (neutrinos, etc.) emissions that are detected by many space and ground telescopes 
in operation. Such detectors, sensitive even in the very high energy regime are mentioned: the KM3NeT, 
IceCube, ANTARES, etc. for neutrinos \cite{Aartsen} and the C.T.A., MAGIC, H.E.S.S, Fermi-LAT, etc. 
for gamma-rays \cite{Ahnen}.

Among the various binary systems, the X-ray binaries (XRBs) consist of two star components being in 
different evolution stage. The compact object (a stellar mass black hole or a neutron star) resulted
from the gravitational collapse of a massive enough that has reached in its final stage. At the
equatorial region of the compact star, mass is accreting out of its companion star. Thus, an accretion disk of matter and gas of extreme temperatures is formed around the compact 
component of XRBs. Moreover, relativistic magnetohydrodynamical flows, collimated and accelerated by magnetic 
fields, are being ejected perpendicularly to the accretion disk. Eventually, the binary system constitutes a prominent source of X-ray, but also $\gamma$-ray and neutrino emission. 

Concerning this work, the model employed \cite{Romero,Reynoso,Reynoso-2019} considers hadronic (with ratio 
$\alpha =L_{p}/L_{e}$) conic jets with half-opening angle $\xi$ and radius $r(z) = z \tan{\xi}$.  
The energy rate transferred to the jet is only 10\% of the system's Eddington luminosity while 
the magnetic field responsible for the jet collimation and acceleration is given by the 
equipartition of magnetic and kinetic energy density as $B=\sqrt{8\pi\rho _k(z)}$ (see 
Ref. \cite{Romero,Reynoso,Reynoso-2019}). 

The acceleration process includes a small portion of the jet protons declared as  
$q_{rel}$ to be further accelerated within a jet region from distances $z_0$ to $z_{max}$. 
The respective acceleration rate is approximated by $t_{acc}^{-1}\simeq\eta ceB/E_p$, with $\eta $
denoting the acceleration efficiency, and is a consequence of shock-waves and the 2nd order Fermi 
mechanism. The velocities acquired belong to nearly relativistic regime. The proton distribution 
describing the process is a power-law given in the jet's rest frame as $N'(E')=K_0E'^{-2}$ $GeV^{-1}cm^{-3}$,
where $K_0$ is a normalization constant. Several cooling mechanisms such as syncrotron emission, 
collisions with the rest of the jet matter and adiabatic jet expansion tend to stall the proton 
energetic boosting and set an upper limit for particle energy usually of the order $E_{max}\approx 10^7$ GeV 
for protons and secondary particles \cite{Papav-Papad-Kosm,Papad-Papav-Kosm}. In previous works, 
we used this model for the neutrino and gamma-ray emission prediction from extragalactic system 
LMC X-1 \cite{Papavasileiou2021}. 

Moreover, another mechanism leading to secondary particle creation and photon emission is relativistic 
proton interactions with photons created inside or outside the jet region. This mechanism was the center 
of a previous work \cite{Papavasileiou2022}, but it is not included in this paper. 

\section{Description of the theoretical method and the simulation procedure }

Several theoretical models have been developed in an attempt to predict and explain the observed 
spectra of mostly Galactic binaries \cite{Aartsen,Ahnen} (see further details see 
\cite{Papavasileiou2021, Papad-Ody-AHEP}). 
In this work, we employ a lepto-hadronic model and consider the injection of both electrons 
and protons at different positions in the jet's base. Then we compute the broad-band photon 
(and neutrino) emission as well as photon absorption by the synchrotron process as well as by 
their interactions with ambient photons and matter. We pay special attention on correcting the 
calculated high energy $\gamma$-ray photon flux by the effect of p-$\gamma$ and $\gamma\gamma$ 
absorption. In this way the surviving flux that arrives on Earth is obtained which can be 
compared with the observed fluxes by the gamma-ray detectors mentioned before.

The basic equation that characterizes the lepto-hadronic model is the transport equation which
describes the concentration (distribution) of particles (mainly proton, pions and muons)
$N_j(E,{\bf r},t)$, where $j = p, e^\pm, \pi^\pm, \mu^\pm$, etc., as a function of the time $t$, 
the particle's energy $E$, and the position ${\bf r}$ inside the jet which is consider of conical
shape.

\subsection{The transport equation for particles moving inside jets} 

In essence, the general form of the transport equation is a phenomenological macroscopic 
integro-differential equation describing the transport (propagation) of particles (or radiation) 
inside the astrophysical outflows (jets) written as
\begin{equation}
\frac{\partial N_j}{\partial t} - \nabla\cdot\left( D_j\nabla N_j\right) + 
\frac{\partial\left(b_j N_j\right)}{\partial E}-\frac{1}{2}\frac{\partial^2\left(d_j N_j\right)}{\partial E^2}
= Q_j(E,{\bf r},t) - p_jN_j + \sum_k \int P_j^k (E^\prime,E) N_j(E,{\bf r},t) dE \, , 
\label{Gen_Trans_Equ}
\end{equation}
\noindent
where $j = p, \pi^\pm, \mu^\pm \, ...$. The parameters $D_j$, $b_j$, $d_j$, $p_j$ and $P_j^k$ may depend 
upon the space and time coordinates and also on the energy $E$. Equation (\ref{Gen_Trans_Equ}) coincides 
with the continuity equation for particles of type-$j$, with j=1,2,3 and $(1,2,3)\equiv$(p, $\pi$, $\mu$)] 
as stated before. 

The term $Q_j(E,{\bf r},t)$ in the r.h.s. of Eq. (\ref{Gen_Trans_Equ}) is equal to the intensity of 
the source producing the particles-$j$, which is also known as the injection function of the respective particles.
This means that $Q_j(E,{\bf r},t)dEd^3{\bf r}dt$ represents the number of particles kind-$j$ provided
by the reaction sources in a volume element $d^3{\bf r}$, in the energy range between $E$ and $E+dE$
during the time $dt$. In the case when the $j$-type particles are products of a chain reaction (as
it holds assuming the p-p reaction chain described above), the function $Q_j(E,{\bf r},t)$ couples 
the $j$-reaction with its parent reaction.

For simplicity, the numerical calculations are performed by assuming a steady-state transfer 
equation for the distributions of particles which considers only the energy-altering mechanisms 
along with particle decay and escape from the jet. The corresponding solution that gives the 
respective particle energy distributions is written as 
\begin{equation}
N(E,z)=\frac{1}{\mid b(E) \mid}\int_{E}^{E_{max}} Q(E',z)e^{-\tau (E,E')}dE' , 
\qquad \tau (E,E')=\int_{E}^{E'} (dE"t^{-1})/\mid b(E")\mid .
\end{equation}
In the solution presented, $N(E,z)$ is the particle number per unit of energy and volume ($GeV^{-1}cm^{-3}$) 
while $Q(E,z)$ is the injection function that translates to the particle production rate. Also, 
$t^{-1}=t_{esc}^{-1}+t_{dec}^{-1}$ denotes the particle reduction rate within the jet. Finally, the 
energy loss rate, $b(E)=-Et_{loss}^{-1}$, introduces all the cooling mechanisms that affect particle 
energy inside the jet such as sunchrotron emission, collisions with jet cold matter and jet adiabatic 
expansion. The proton injection function in the observer's reference frame is given by
\begin{align}
\label{eq6}
Q_{p}(E_p,z)&=Q_0\left(\frac{z_0}{z}\right)^3
\Gamma_b^{-1} \left(E_p -\beta_b \cos i\sqrt{E_p^2-m^2c^4}\right)^{-2}
\left[\sin^2 i + \Gamma_b^2\left(\cos i -\frac{\beta_bE_p}{\sqrt{E_p^2-m^2c^4}}\right)^2\right]
^{-1/2} .  
\end{align}
In the above transformation, $\Gamma _b$ corresponds to the Lorentz factor of the jet's bulk velocity $\upsilon _{b}=\beta_{b}c$ and $i$ to the inclination of the system (i.e, the angle to the line of sight). $Q_{0}$ is the normalization constant that can be found in \cite{Romero} or \cite{Papavasileiou2021}. The cold proton density enters the calculations as $n(z)= (1-q_{rel})L_k/(\Gamma _{b} m_pc^2\pi r(z)^2\upsilon _{b})$.

We need the charged pion spectrum per $p-p$ collision in order to calculate the respective distribution. This is given in \cite{Kelner} as
\begin{align}
F_{\pi}\left(x,\frac{E}{x}\right)&=4\alpha B_{\pi}x^{\alpha -1}\left(\frac{1-x^{\alpha}}{1+rx^{\alpha}(1-x^{\alpha})}\right)^4 
\left(\frac{1}{1-x^{\alpha}}+\frac{r(1-2x^{\alpha})}{1+rx^{\alpha}(1-x^{\alpha})}\right)\left(1-\frac{m_{\pi}c^2}{xE_p}\right)^{1/2} ,
\end{align}
where $x=E_{\pi}/E_{p}$, $B_{\pi}=\alpha '+0.25$, $\alpha '=3.67+0.83L+0.075L^2$, $r=2.6/\sqrt{\alpha '}$, and $\alpha=0.98/\sqrt{\alpha '}$. Also, it is $L=ln(E_p/1000$ $GeV)$. The muon spectra from pion decay are written as \cite{Lipari} 
\begin{equation}
\mathcal{N} _{\mu}^{+}=\frac{r_{\pi}(1-x)}{E_{\pi}x(1-r_{\pi})^2}\Theta (x-r_{\pi}) \, , \quad 
\mathcal{N} _{\mu}^{-}=\frac{(x-r_{\pi})}{E_{\pi}x(1-r_{\pi})^2}\Theta (x-r_{\pi}) ,
\end{equation}
where $x=E_{\mu}/E_{\pi}$, $r_\pi =(m_\mu/m_\pi)^2$ and $\Theta$ the Heaviside function. In addition, the respective neutrino emissivity from pion decay is the following
\begin{align}
Q_{\nu}^{(\pi)}(E_{\nu},z)&=\int_E^{E_{max}}t_{\pi,dec}^{-1}(E_\pi) N_\pi(E_\pi,z)\frac{\Theta 
(1-r_\pi-x)}{E_\pi(1-r_\pi)} dE_\pi ,
\end{align}
where $x=E_{\nu}/E_{\pi}$ and the pion decay rate is given by $t_{\pi , dec}^{-1}=(2.6\times 10^{-8}\gamma _{\pi})^{-1}$ $s^{-1}$.

In order to calculate the produced gamma-ray emissivity, we need the gamma-ray spectrum per $p-p$ collision that includes $\pi ^{0}$ and $\eta$ decays \cite{Kelner}
\begin{align}
F_{\gamma}(x,E_p)=B_{\gamma}\frac{lnx}{x}\left(\frac{1-x^{\beta_{\gamma}}}{1+k_{\gamma}x^{\beta_{\gamma}}(1-x^{\beta_{\gamma}}}\right)^4
\left(\frac{1}{lnx}-\frac{4\beta_{\gamma}x^{\beta_{\gamma}}}{1-x^{\beta_{\gamma}}}-\frac{4k_{\gamma}\beta_{\gamma}x^{\beta_{\gamma}}(1-2x^{\beta_{\gamma}})}{1+k_{\gamma}x^{\beta_{\gamma}}(1-x^{\beta_{\gamma}})}\right) ,
\label{Gamma-ray spectra}
\end{align}
where $x=E_{\gamma}/E_{p}$, $B_{\gamma}=1.3+0.14L+0.011L^2$, $\beta_{\gamma}=1/(0.008L^2+0.11L+1.79)$, and $k_{\gamma}=1/(0.014L^2+0.049L+0.801)$. These results are consistent with proton energies in the range $100$ $GeV<E_p<10^8$ $GeV$.

\section{Gamma-ray annihilation}

There are plenty of ambient photon fields capable of absorbing the jet-emitted high-energy radiation that would otherwise be heading towards the Earth. Some of the most important are the X-ray emission from the system's accretion disk and the donor star's UV thermal emission.

\subsection{Accretion disk}

The geometry of disk and analysis we adopt is described in \cite{Cerutti}. Thus, we consider an horizontal and geometrically thin disk with high optical thickness that begins from $R_{in}=6R_{g}$ (i.e, the ISCO-radius for a non-spinning black hole). Every surface element is at thermal equilibrium so the emitted spectrum is given by a black body distribution as
\begin{equation}
\frac{dn}{d\epsilon d\Omega}= \frac{2}{h^{3}c^{3}}\frac{\epsilon ^{2}}{e^{\mathcal{T}(R)}-1},
\end{equation}   
where $\mathcal{T}(R)= \epsilon /k_{B}T(R)$. The temperature profile and its radial dependence is chosen as to respect the disk boundary conditions as (see Ref. \cite{Cerutti}) 
\begin{equation}
T(\bar{r})= T_{g}\left[\frac{\bar{r}-2/3}{\bar{r}(\bar{r}-2)^{3}}\left(1-\frac{3^{3/2}(\bar{r}-2)}{2^{1/2}\bar{r}^{3/2}}\right)\right]^{1/4},
\end{equation}  
where the radius $\bar{r}= R/R_{g}$ is dimensionless. The temperature in the inner zones of the disk is the following
\begin{equation}
\label{Temp_g}
T_{g}= \left(\frac{3GM_{BH}\dot{M}_{accr}}{8\pi \sigma _{SB}R_{g}^{3}}\right)^{1/4},
\end{equation} 
where $\sigma _{SB}$ the Stefan-Boltzmann constant, $R_{g}$ the gravitational radius and $\dot{M}_{accr}$ the system's accretion rate that in most cases corresponds to the Eddington limit $L_{disk}\approx 10^{38}$ $erg/s$.

\subsection{Donor star}

The photon field density of the donor star's thermal emission is given by
\begin{equation}
\frac{dn_{ph}}{d\epsilon}= \frac{15}{4\pi ^{5}c}\frac{L_{star}\mathcal{T}^{4}}{\epsilon ^{2}\ \left(e^{\mathcal{T}}-1\right)},
\end{equation}
where the dimensionless $\mathcal{T}= \epsilon /k_{B}T_{eff}$. $T_{eff}$ represents the effective stellar surface temperature. It is defined by the respective stellar luminosity and radius. The geometry and analysis implemented for the optical depth calculation was completed in \cite{Bottcher}. An important system parameter here is the binary separation $s$, the distance between the compact object and the donor star. For our calculations, we consider that $s=10^{12}$ $cm$ which corresponds to an order of magnitude that holds for many X-ray binary systems.

\begin{table}
\caption{\label{Table1} Model parameters in numeric calculations.}
\begin{ruledtabular}
\begin{center}
\begin{tabular}{l l l l l l l l l}
Parameter & Value & Unit & Parameter & Value & Unit & Parameter & Value & Unit \\ [0.2ex]
\hline \\[0.2ex]
$q_{rel}$ & $10^{-4}$ & - & $M_{BH}$ & $15$ & \(M_\odot\) & $s$ & $10^{12}$ & $cm$ \\ [0.2ex]
$\eta$ & $0.1$ & - & $M_{donor}$ & $20$ & \(M_\odot\) & $\upsilon _{b}$ & $0.7c$ & - \\ [0.2ex]
$\alpha$ & $100$ & - & $\dot{M}_{accr}$ & $10^{-8}$ & \(M_\odot\)/yr & $\xi$ & 2 & $^\circ$ \\ [0.2ex]
$z_{0}$ & $6.34\times 10^{8}$ & $cm$ & $i$ & 30, 70 & $^\circ$ & $d$ & $2$ & $kpc$ \\ [0.2ex]
$z_{max}/z_{0}$ & $5$ & - & $L_{star}$ & $10^{5}$ & \(L_\odot\) & $P_{orb}$ & $15$ & $days$ \\ [0.2ex]
$R_{in}$ & $6R_{g}$ & - & $R_{out}$ & $10^{10}$ & $cm$ & $T_{eff}$ & $10^{4}$ & $K$ \\ [0.2ex] 
\end{tabular}
\end{center}
\end{ruledtabular}
\end{table} 

\section{Results and discussion}

As it is well known, the $p-p$ collisions between relativistic protons and cold protons within 
the jet region result in neutral and charged pion production ($\pi ^{0}$, $\pi ^{\pm}$). The 
first ones decay to gamma-ray photons, $\pi ^{0} \rightarrow\gamma +\gamma $, while the charged 
pions decay into muons and neutrinos as $\pi^+\rightarrow\mu^+ +\nu_{\mu}$ and $\pi^-\rightarrow\mu^-
+\bar{\nu}_{\mu}$
Also, the $\eta$-mesons decays to gamma-rays, $\eta \rightarrow\gamma +\gamma$, are included to 
gamma-ray spectrum in the calculations. The emitted high-energy photons, though, are annihilated 
by low-energy ambient photons as $\gamma +\gamma\rightarrow e^+ + e^-$.

In this work, we use the system and model parameters of Table \ref{Table1} that refer to a standard, 
average Black hole X-ray binary system to calculate the particle distributions and emissivities of 
the previous sections.

\begin{figure*}[ht] 
\begin{tikzpicture}
\centering
  \node (img1)  {\includegraphics[width=0.5\linewidth]{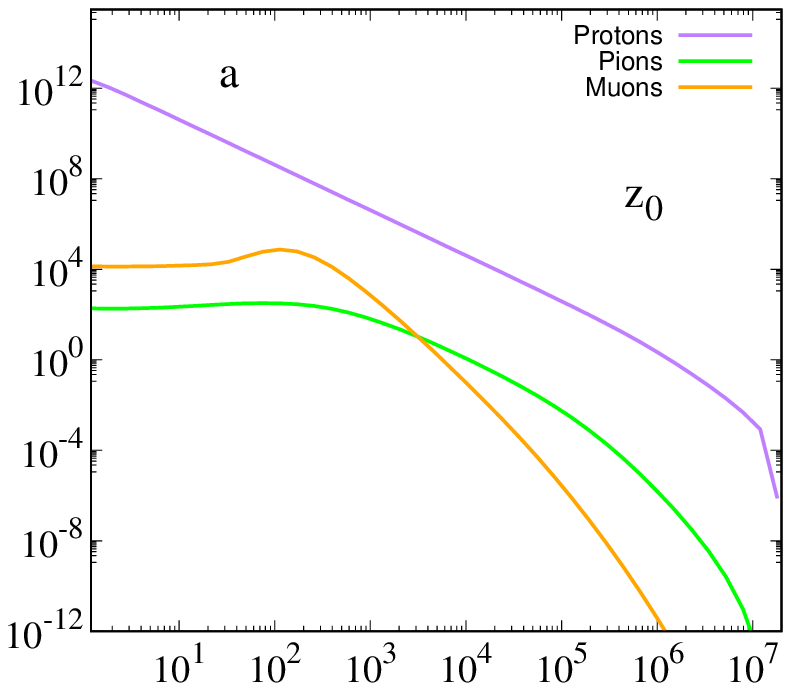}};
  \node[left= of img1, node distance=0cm, rotate=90, anchor=center,yshift=-1.2cm, font=\color{black}, font=\normalsize] {$N$ [$GeV^{-1}cm^{-3}$]};
  \node[below= of img1, node distance=0cm, yshift=1cm, xshift=0.5cm, font=\color{black}, font=\normalsize] {E [GeV]};
  \node[right= of img1, xshift=-1.2cm] (img2)  {\includegraphics[width=0.5\linewidth]{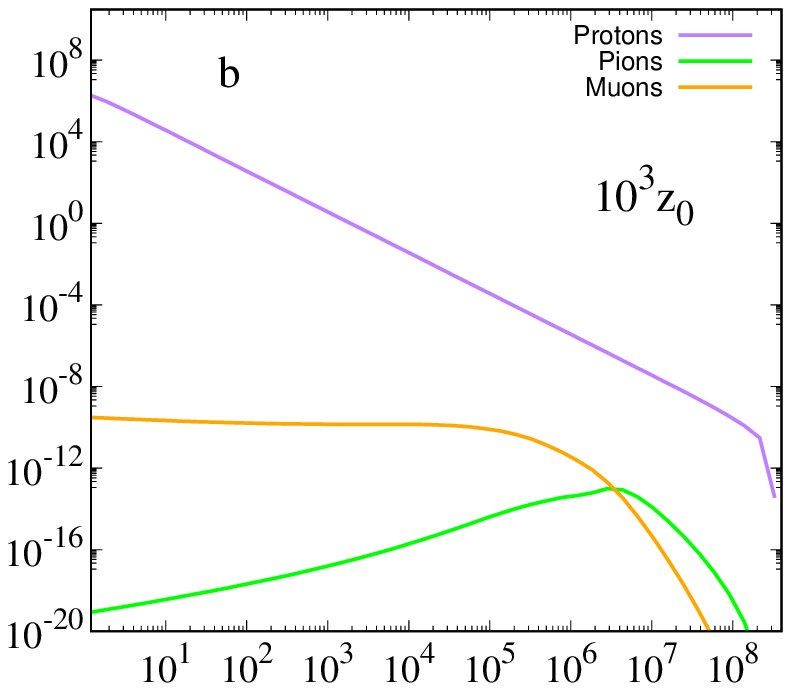}};
  \node[below= of img2, node distance=0cm, yshift=1cm, xshift=0.5cm, font=\color{black}, font=\normalsize] {E [GeV]};
\end{tikzpicture}

\caption{\label{figure1} Proton, pion and muon energy distributions for (\textbf{a}) $z_{0}$ and (\textbf{b}) $10^{3}z_{0}$. }
\end{figure*}

In Fig. \ref{figure1}, the ralativistic proton, pion and muon densities per unit of energy produced within the jet region dictated by the boundaries of Table \ref{Table1} are presented. Each graph refers to the labeled distance from the compact object, $z_{0}$ and $10^{3}z_{0}$. The reason for muon dominance over pions for $E<10^{4}$ $GeV$ (graph \textbf{a}) is the the far greater pion decay rate compared to muons. On the other hand, superior synchrotron radiation rate (due to smaller mass) is to be blamed for the steeper decline of muon distribution instead. In graph (\textbf{b}), the jet expansion with the distance has catastrophic results in pion and, therefore, muon production efficiency. This is attributed mainly to the drastic increase of $p-p$ collision mean free path. That is why pion and muon reduction is much greater than expected from proton comparisons between graphs (\textbf{a}) and (\textbf{b}). Moreover, a better collimated jet that is differentiated from the basic conical shape should be able to maintain high efficiency in particle production over higher jet regions. 

\begin{figure*}[ht] 
\begin{tikzpicture}
\centering
  \node (img1)  {\includegraphics[width=0.5\linewidth]{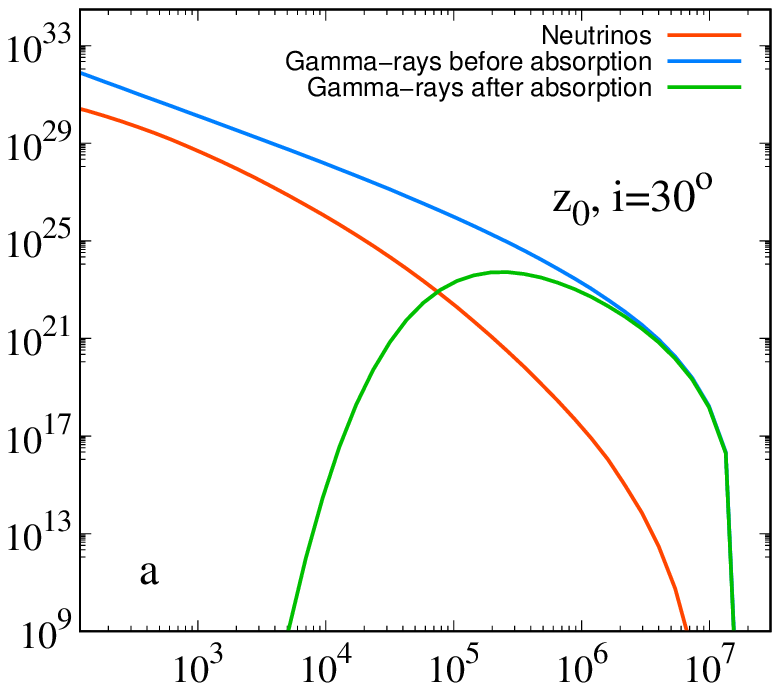}};
  \node[left= of img1, node distance=0cm, rotate=90, anchor=center,yshift=-1.2cm, font=\color{black}, font=\normalsize] {$I$ [$GeV^{-1}s^{-1}$]};
  \node[right= of img1, xshift=-1.2cm] (img2)  {\includegraphics[width=0.5\linewidth]{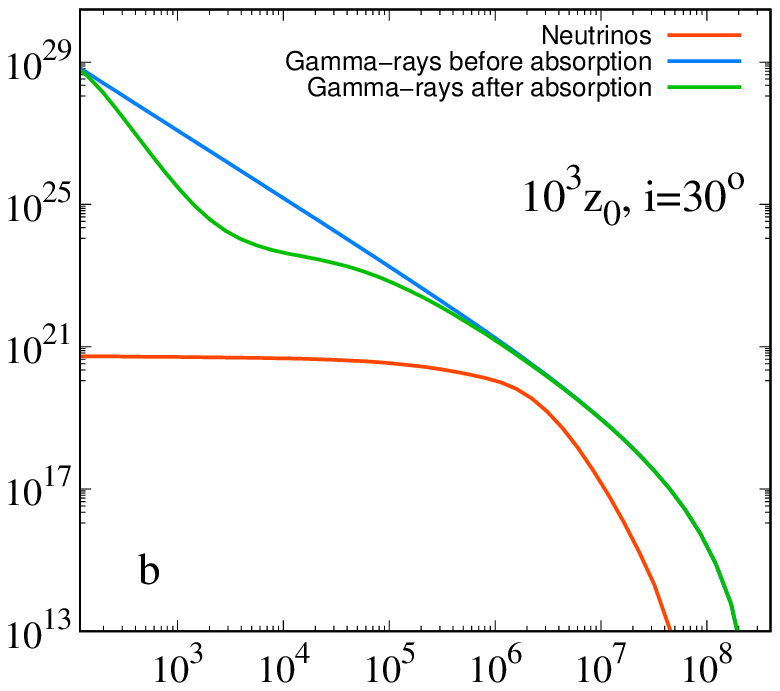}};
  \node[below= of img1, yshift=1cm] (img3)  {\includegraphics[width=0.5\linewidth]{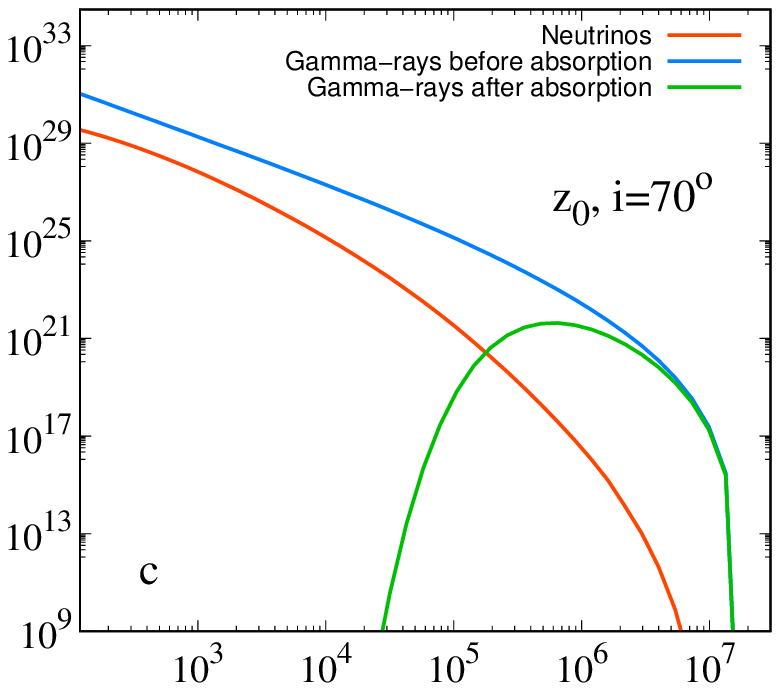}};
  \node[left= of img3, node distance=0cm, rotate=90, anchor=center,yshift=-1.2cm, font=\color{black}, font=\normalsize] {$I$ [$GeV^{-1}s^{-1}$]};
  \node[below= of img3, node distance=0cm, yshift=1cm, xshift=0.5cm, font=\color{black}, font=\normalsize] {E [GeV]};
  \node[right= of img3, xshift=-1.2cm] (img4)  {\includegraphics[width=0.5\linewidth]{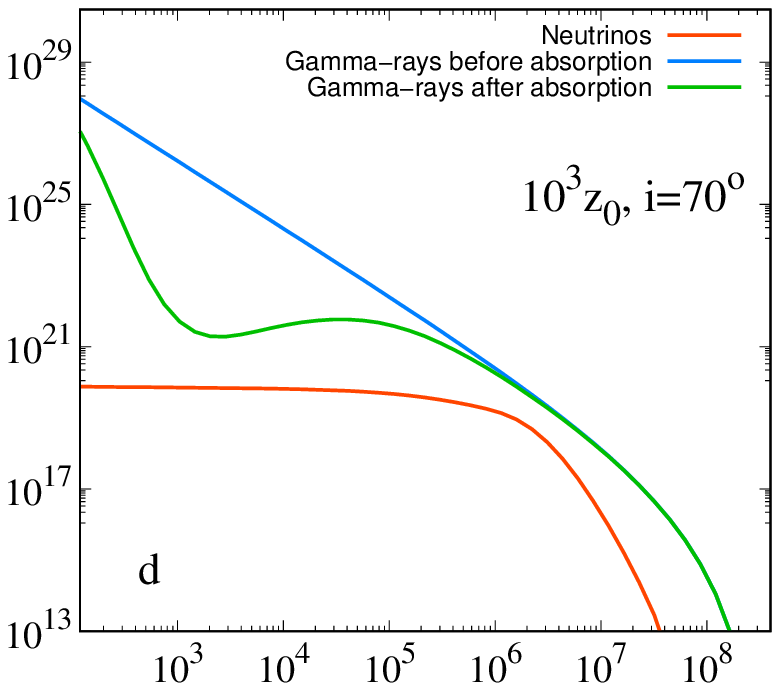}};
  \node[below= of img4, node distance=0cm, yshift=1cm, xshift=0.5cm, font=\color{black}, font=\normalsize] {E [GeV]};
\end{tikzpicture}

\caption{\label{figure2} Neutrino and gamma-ray produced intensities from $\pi ^{\pm , 0}$ and $\eta$-meson decays within the jet. The top panel (\textbf{a}, \textbf{b}) corresponds to system inclination of $i=30^\circ$ but different jet-source distances ($z_{0}$ and $10^{3}z_{0}$) from the black hole while the bottom panel (\textbf{c}, \textbf{d}) concerns inclination of $i=70^\circ$.}
\end{figure*}

Concerning Fig .\ref{figure2}, neutrino intensity tends to imitate the behavior of muon distribution. Obviously, both muons and neutrinos in this case are products of pion decay with similar restrictions imposed by the step function. We notice the clear decline of neutrino production in graphs (\textbf{b}) and (\textbf{d}) compared to (\textbf{a}) and (\textbf{c}). As mentioned, the reason is found in the growing insufficiency of secondary particle production with the distance from the center.

In addition to neutrinos, in Fig. \ref{figure2} we also present the respective gamma-ray intensities integrated over the acceleration jet regions that start from $z_{0}$ (graphs \textbf{a} and \textbf{c}) and $10^{3}z_{0}$ (graphs \textbf{b} and \textbf{d}). We exhibit the intensity curves before and after absorption due to the accretion disk's X-ray emission and donor star's thermal one. In lower jet regions, the disk is the supreme cause of absorption while for higher jet parts it loses that role to the companion star. Decreasing the angle to the sight of line (i.e, from $i=70^\circ$ to $30^\circ$) impacts positively but not overly every intensity curve, including neutrinos. On the other hand, absorption grows weaker cementing the higher detection probabilities.         

\section{Summary and Conclusions}
 
Relativistic astrophysical jets consist of accelerated leptonic 
and/or hadronic content. Various mechanisms such as shock-waves are 
able to accelerate a portion of those even further the relativistic scale 
causing collisions with the rest of the jet matter and secondary particle production.
We are interested in $p-p$ interaction mechanism that results in pion, muon, neutrino 
and gamma-ray production, the latter of those are emitted from the jets towards even the Earth. 
However, important obstacle in gamma-ray detection is the absorption due to lower-energy emissions from the system's disk and companion star. 

We employ the steady-state transfer equation, the corresponding particle injection functions and energy-loss mechanisms in order to calculate the produced relativistic proton, pion and muon energy distributions. We find the pion production efficiency reduced in greater distances inside the jet and, thus, deem the conic jet geometry rather inappropriate to maintain the secondary particle production rate in higher jet regions. Also, we demonstrate the marginally higher neutrino and gamma-ray intensities combined with weaker gamma-ray absorption for smaller system inclinations.



\end{document}